\documentclass[prl,aps,epsfig,superscriptaddress]{revtex4}
\usepackage{bm}
\usepackage{amsfonts}
\usepackage[dvips]{graphicx}
\usepackage{mathrsfs}
\usepackage[intlimits]{amsmath}
\usepackage[colorlinks, citecolor=red]{hyperref}

\begin{document}

\title{Experimental realization of $105$-qubit random access quantum memory}
\author{N. Jiang$^{1}$\footnote{%
These authors contributed equally to this work.\label{FN}}, Y.-F. Pu$^{1}$\footnotemark[1]\footnote{Present address: Institute for Experimental Physics, University of Innsbruck, A-6020 Innsbruck, Austria}, W. Chang$^{1}$, C. Li$^{1}$%
, S. Zhang$^{1}$, L.-M. Duan\footnote{Corresponding author: lmduan@tsinghua.edu.cn}}
\affiliation{Center for Quantum Information, IIIS, Tsinghua University, Beijing 100084,
PR China}
%\affiliation{Department of Physics, University of Michigan, Ann Arbor, Michigan 48109, USA}

\begin{abstract}
Random access memory is an indispensable device for classical
information technology. Analog to this, for quantum information technology,
it is desirable to have a random access quantum memory with many memory
cells and programmable access to each cell. We
report an experiment that realizes a random access quantum memory of
$105$ qubits carried by $210$ memory cells in a macroscopic atomic ensemble.
We demonstrate storage of optical qubits into these memory cells and their read-out at programmable times by
arbitrary orders with fidelities exceeding any classical bound. Experimental
realization of a random access quantum memory with many memory cells and
programmable control of its write-in and read-out makes an important step
for its application in quantum communication, networking, and computation.
\end{abstract}

\maketitle

\section{Introduction}

Classical random access memory, with its programmable access to many memory
cells and site-independent access time, has found wide applications in
information technologies. Similarly, for realization of quantum
computational or communicational networks, it is desirable to have a random
access quantum memory (RAQM) with the capability of storing many qubits,
individual addressing of each qubit in the memory cell, and programmable
write-in and read-out of the qubit from the memory cell to a flying bus
qubit with site-independent access time \textbf{\cite{1,2,3,4,5,6,7,8,9,10}}.
Such a device is useful for both
quantum communication and computation \cite{1,2,3,4,5,6,7,8,9,10}. It
provides a key element for realization of long-distance quantum
communication through the quantum repeater network \cite{1,2,3,4,5,6,7}.
The write-in and read-out operations require implementation of a good
quantum interface between the bus qubits, typically carried by the photonic
pulses, and the memory qubits, which are usually realized with the atomic
spin states. A good quantum interface should be able to faithfully map
quantum states between the memory qubits and the bus qubits. A convenient implementation of the quantum interface is based
on the directional coupling of an ensemble of atoms with the forward
propagating signal photon pulse induced by the collective enhancement effect
\textbf{\cite{2,3,4,5,6,10a}}. A number of experiments have demonstrated
this kind of quantum interfaces and their applications both in the atomic
ensemble \textbf{\cite{10a,11,12,13,14,15,16,18,19,20,20a,31,32}} and the
solid-state spin ensemble with a low-temperature crystal \textbf{\cite%
{21,22,23,23a,24,25}}.

To scale up the capability of a quantum memory, which is important for its
application, an efficient method is to use the memory multiplexing, based on
the use of multiple spatial modes \cite{16,20}, or temporal modes \textbf{%
\cite{22,23,24,25}}, or angular directions \cite{20a} within a single atomic
or solid-state ensemble. Through multiplexing of spatial modes, recent
experiments have realized a dozen to hundreds of memory cells in a single
atomic ensemble, however, write-in and read-out of external quantum signals
have not been demonstrated yet \cite{16,20}. With temporal multiplexing, a
sequence of time-bin qubits have been stored into a solid-state ensemble,
however, the whole pulse sequence needs to be read out together with a fixed
order and interval between the pulses for lack of individual addressing
\textbf{\cite{23,24,25}}. It remains a challenge to demonstrate a RAQM with
programmable and on-demand control to write-in and read-out of each
individual quantum signals stored into the memory cells.

In this paper, we demonstrate a RAQM which can store $105$ qubits in its $210
$ memory cells using the dual-rail representation of a qubit. A pair of
memory cells stores the state of the input photonic qubit, which is carried
by the two paths of a very weak coherent pulse. We have measured the
fidelities and the efficiencies for the write-in, storage, and readout
operations for all the $105$ pairs of memory cells. The fidelities,
typically around or above $90\%$, are significantly higher than the
classical bound and therefore confirm quantum storage. To demonstrate the
key random access property, we show that different external optical qubits
can be written into the multi-cell quantum memory, stored there
simultaneously, and read out later on-demand by any desired order with the
storage time individually controlled for each qubit. The fidelities for all
the qubits still significantly exceed the classical bound with negligible
crosstalk errors between them.

\section{Results}
\subsection{Experimental setup}

Our experimental setup is illustrated in Fig. 1. The macroscopic atomic
ensemble is realized with a cloud of $^{87}$Rb atoms trapped and cooled down
by a magneto-optical trap (MOT) (see Supplementary Note 1). We divide this
macroscopic ensemble into a two-dimensional (2D) array of $15\times 14$
micro-ensembles. Each micro-ensemble is individually addressed through a
pair of crossed acoustic optical deflectors (AODs) inserted into the paths
of the control beam, the input probe beam, and the output probe beam (see
Supplementary Note 2). The AODs provide a convenient device for multiplexing
and de-multiplexing of many different optical paths, which have been used
recently to control neutral atoms \cite{16,20,27b} as well as trapped ions \cite%
{27}. The relative phases between those $210$ different optical paths are
intrinsically stable as the beams along different paths go through the same
optical devices.

\begin{figure}[tbp]
\includegraphics[width=16cm]{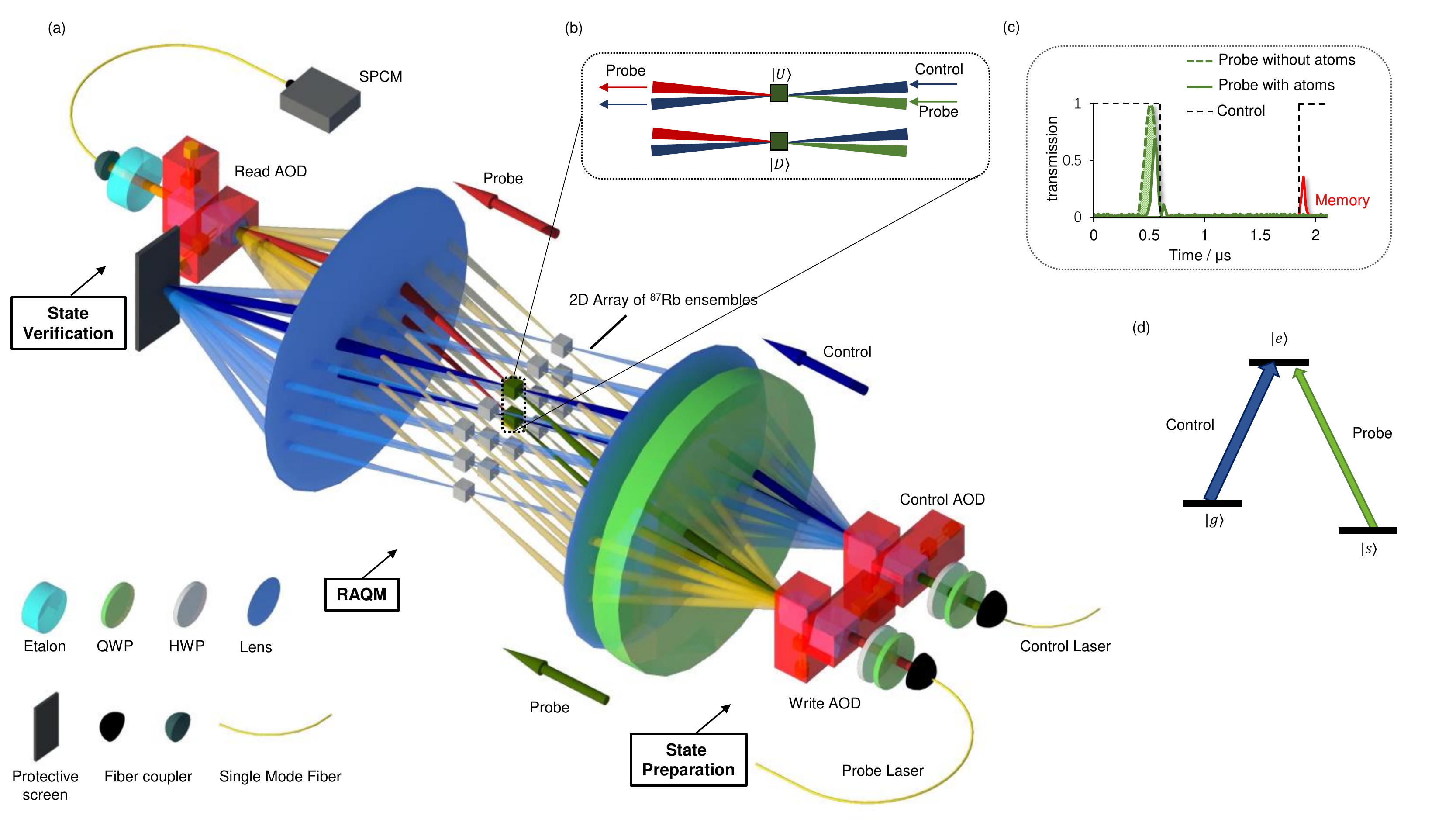} %
\caption{\textbf{Experimental setup for realization of a multiplexed random access
quantum memory with a macroscopic atomic ensemble. }\textbf{(a)} We use two-dimensional (2D)
AODs to control the multiplexing of control and probe beams which are coupled to a macroscopic
atomic ensemble under the EIT configuration. The qubit state of the probe photon is stored into
a pair of memory cells using the dual-rail representation and later retrieved for read out after a controllable storage time. Through
programming the AODs, we individually address and control a 2D array of $15 \times 14$ atomic memory
cells (for clarity only $3 \times 5$ cells are shown in the figure), which can store $105$ optical qubits.
The lens, set in a $4f$-configuration, are used to focus the beams as well as to map different
angles of the deflected beams after the AODs to different micro-ensembles in the 2D array.
We use a Fabry-Perot cavity (etalon) in
the path of the retrieved photon for frequency filtering of the leaked control light. The write AOD prepares the input state for the atomic memory,
where the optical qubit is carried by superposition of different optical paths. This input state, unknown to the atomic memory, is stored into the multi-cell atomic ensemble. After a controllable storage time, this state is mapped out to the optical state carried by different optical paths (on the left side of the atomic ensemble) and verified in complementary qubit bases by a combination of the read AOD and the SPCM (single-photon counting module). \textbf{(b)} Zoom-in of the
beam configuration at two memory cells denoted as U and D for qubit storage.
\textbf{(c)} Histogram of the time-resolved photon counts for the transmitted  probe light registered by the single-photon detector.
The solid (dashed) green curves represent respectively the probe pulse with (without) the MOT atoms, and
their difference corresponds to the stored photon component. The red curve
represents the retrieved photon pulse after a controllable storage time. \textbf{(d)} The energy levels
of the ${}^{87}Rb$ atoms coupled to the control and the probe beams through the EIT configuration, with $|g\rangle \equiv
|5S_{1/2},F=2\rangle $, $|s\rangle \equiv |5S_{1/2},F=1\rangle $, $|e\rangle
\equiv |5P_{1/2},F^{\prime }=2\rangle $.}%

\end{figure}

The atoms in the whole ensemble are initially prepared to the state $%
|g\rangle \equiv |5S_{1/2},F=1\rangle $ through optical pumping and the MOT\
is turned off right before the quantum memory experiment. For each
micro-ensemble, the probe and the control beams are interacting through the
electromagnetically-induced transparency (EIT) configuration shown in Fig. 1
\textbf{\cite{10a}}, where an incoming photon in the probe beam is converted
by the control beam (the write pulse) to a collective spin wave excitation
in the level $|s\rangle \equiv |5S_{1/2},F=2\rangle $ through the excited
state $|e\rangle \equiv |5P_{1/2},F^{\prime }=2\rangle $. The write pulse is
then shut off. After a programmable storage time in the quantum memory, the
excitation in the spin wave mode is converted back to an optical excitation
in the output probe beam by shining another pulse (the read pulse) along the
control beam direction.

\subsection{Characterization of storage fidelities for every memory cells}

The input qubit state is carried by two optical paths $\left\vert
U\right\rangle $ and $\left\vert D\right\rangle $ of a photon, and any
superposition state $c_{0}\left\vert U\right\rangle +c_{1}\left\vert
D\right\rangle $ with arbitrary coefficients $c_{0},c_{1}$ can be generated
and controlled through the input AODs. The input signal is carried by a very
weak coherent pulse with the mean photon number $\bar{n}\simeq 0.5$. In our proof-of-concept experiment,
the input state for the atomic memory is prepared by the write AODs as illustrated in Fig. 1(a).
This pair of AODs can split the weak coherent signal into arbitrary superpositions along two different
optical paths, and the single-photon component of this weak signal represents the effective qubit state,
with the qubit information carried by the superposition coefficients along the different optical paths. Note that the input qubit state,
prepared by the write AODs, remains unknown to the atomic ensemble which acts as the multiplexed quantum memory in this experiment.
This dual-rail encoding of qubit is same as the path or polarization qubit used in many optical quantum information experiments, where
the single-photon component of a very weak coherent state carries the qubit state and is selected out by the single-photon detectors
afterwards.

Similar to other optical quantum information experiments \cite{2,3,4,5,6}, we use two
quantities---conditional fidelity and efficiency---to characterize the imperfections of
a quantum memory. The conditional fidelity characterizes how well the qubit state is preserved when a
photon is registered on the output channel after its storage in the quantum
memory \cite{29b}. The (intrinsic) efficiency characterizes the success probability of a stored
photon reappeared in the output single-mode fiber after a certain storage
time. For application of quantum memory in quantum information protocols, such as for implementation of quantum
repeaters \cite{1,2,3,6}, the conditional fidelity is typically the most important
figure-of-merit as it determines the fidelity of the overall protocol and characterizes whether one can enter the quantum region
by beating the classical bound. The efficiency influences the overall success probability of the quantum information protocol. For
the quantum repeater protocol based on the DLCZ (Duan-Lukin-Cirac-Zoller) scheme \cite{2,3,6}, the scaling of required resources remains polynomial at any finite efficiency, but the scaling exponent gets significantly less when one increases the efficiency. For quantum memory experiments, one
needs to first achieve a high enough conditional fidelity to prove that the system enters the quantum storage region by beating the classical bound and then improve the efficiency as much as one can.

\begin{figure}[ptb]
\includegraphics[width=13cm]{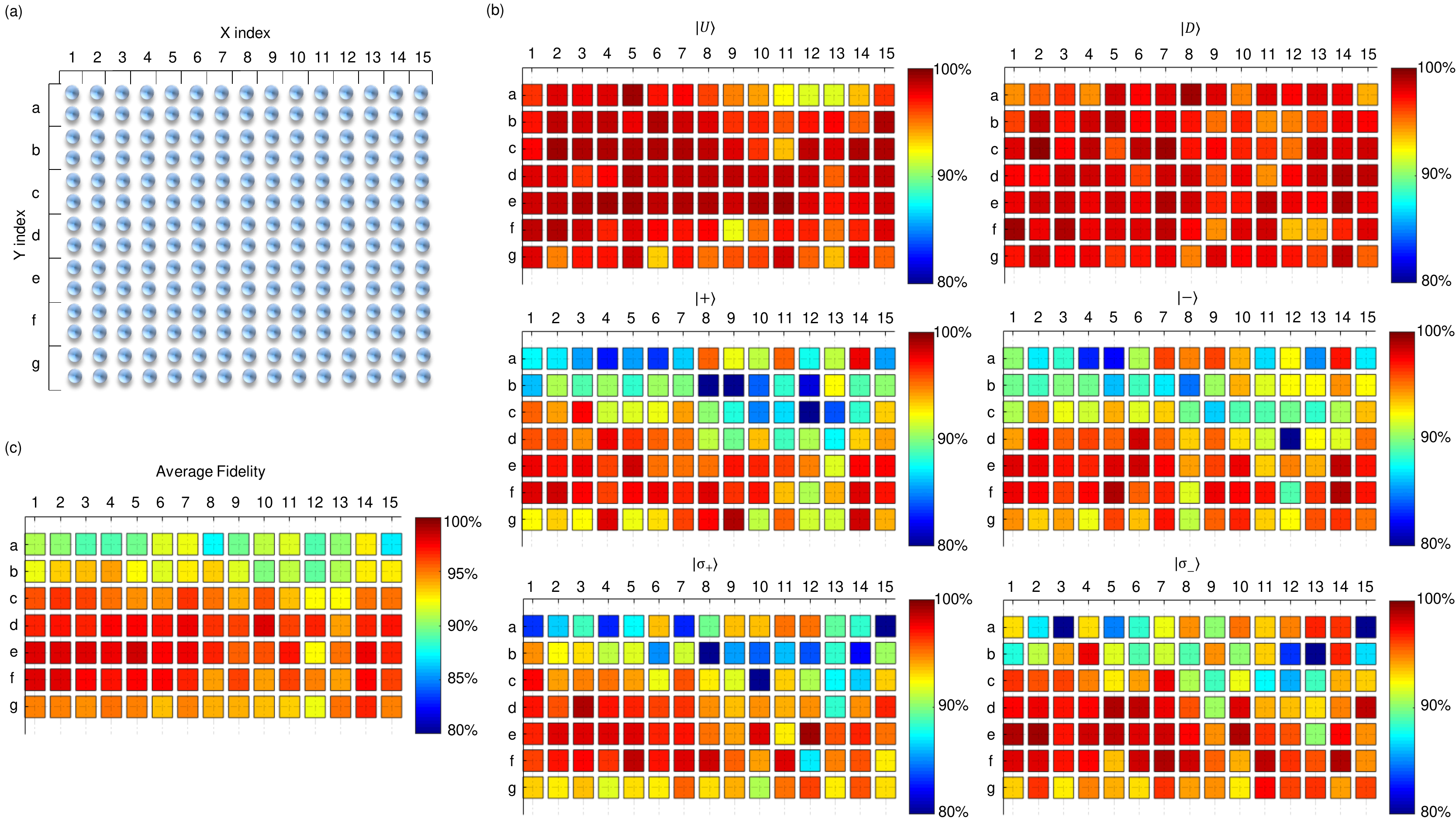}
\caption{\textbf{Measured state fidelities of the retrieved optical qubits after storage in
the $210$-cell quantum memory.} \textbf{(a)} Illustration of the $105$-qubit quantum memory. Each qubit is carried by a pair of neighboring
memory cells in the 2D array. \textbf{(b)} Quantum state fidelities measured for the six complementary input
states of optical qubits after a $1.38$ $\protect\mu s$ storage time. We measured the fidelities for all the
$105$ pairs of memory cells one by one. \textbf{(c)} The average
storage fidelities for the $105$ pairs of memory cells.}
\end{figure}

\begin{figure}[ptb]
\includegraphics[width=14cm]{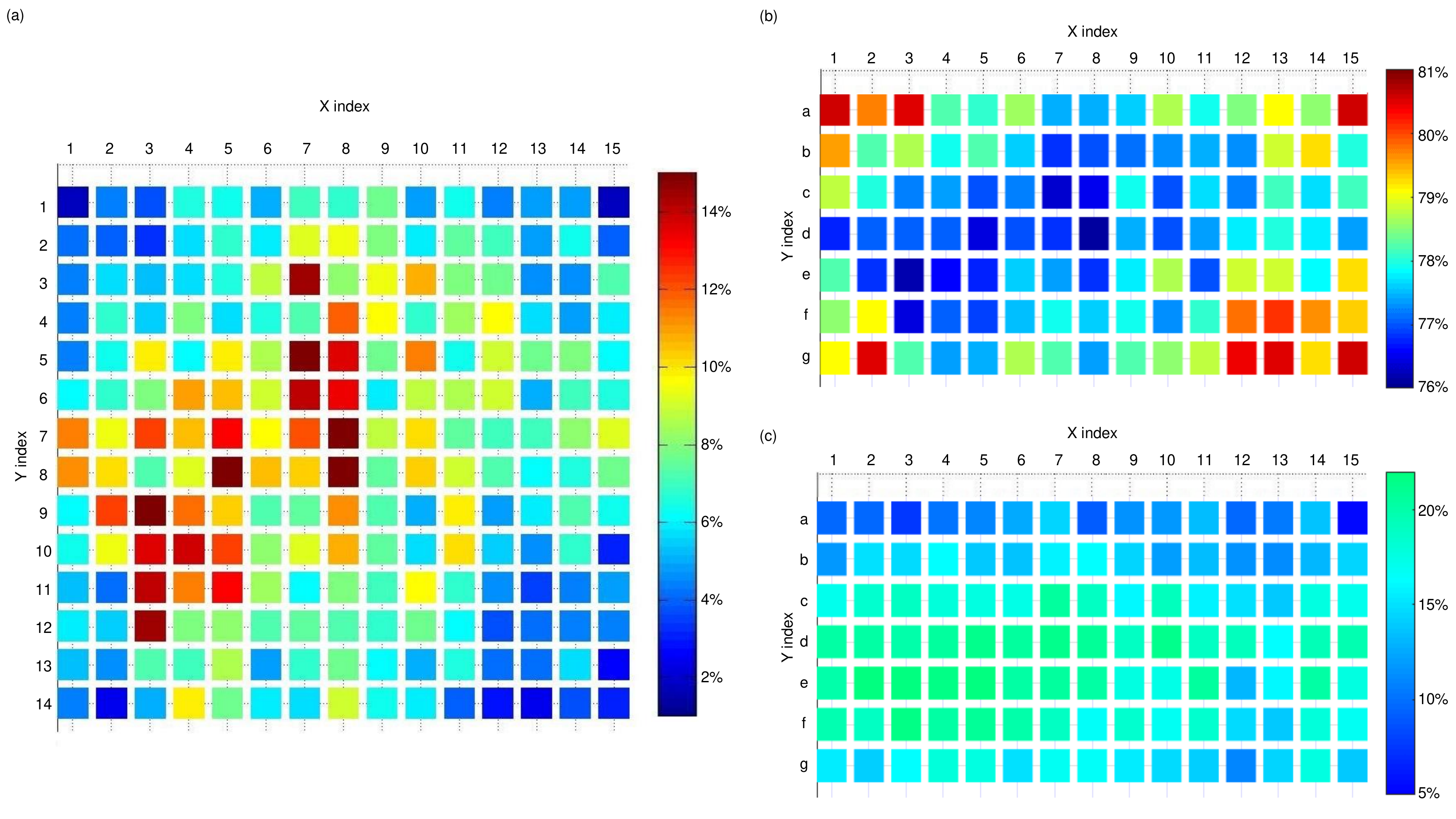}
\caption{ \textbf{Photon retrieval efficiency from the memory cells and its influence on
the classical bound of the storage fidelity.}  \textbf{(a)} Photon retrieval efficiency measured
for the 2D array of $15 \times 14$ memory cells. The storage time here is $%
1.38\protect\mu s$.  \textbf{(b)} The classical bound on the storage fidelity for each pair of
memory cells, taking into account of
the retrieval efficiency and the multi-photon components. \textbf{(c)} The measured storage fidelities subtracted by the
corresponding classical bounds. The positive values indicate that we have demonstrated quantum storage for every pairs of memory cells.}
\end{figure}

%\begin{center}
%{\textbf{The qubit storage in the 105 pairs of cells}}
%\end{center}

The input optical qubit is stored into a pair of neighboring micro-ensembles
through the EIT\ process and then retrieved after a programmable storage
time by controlling the write/read pulses in the corresponding paths. The
write/read pulses are delivered to different paths through the control AODs.
Our atomic quantum memory with a 2D array of $15\times 14$ memory cells has
the capability to store $105$ optical qubits as shown in Fig. 2a. First, we
measure the storage fidelity for each pair of the memory cells one by one by
inputting an optical qubit and then retrieving it for readout after a
storage time of $1.38$ $\mu $s. The output qubit state is measured through
quantum state tomography \cite{30} by using the output AODs to choose the
complementary detection bases. The experimentally reconstructed density
operator $\rho _{o}$ is compared with the input state $\left\vert \Psi
_{in}\right\rangle $ prepared by the input AODs to get the storage fidelity $%
F=\left\langle \Psi _{in}\right\vert \rho _{o}\left\vert \Psi
_{in}\right\rangle $. For each pair of the memory cells, we measure the
storage fidelity $F$ under six complementary input states with $\left\vert
\Psi _{in}\right\rangle $ taking $\left\vert U\right\rangle ,\left\vert
D\right\rangle $, $\left\vert \pm \right\rangle =\left( \left\vert
U\right\rangle \pm \left\vert D\right\rangle \right) /\sqrt{2}$, and $%
\left\vert \sigma _{\pm }\right\rangle =\left( \left\vert U\right\rangle \pm
i\left\vert D\right\rangle \right) /\sqrt{2}$, and the results are shown in
Fig. 2b for all the $105$ pairs of memory cells. The averaged conditional fidelity $%
\overline{F}$, from the above six measurements with equal weight, is shown
in Fig. 2c. For a single-photon input state, the classical bound (maximum
value) for the conditional storage fidelity $\overline{F}$ is $2/3$ (see the Supplement).
When we consider the contribution of small multi-photon components
in the weak coherent pulse (with $\bar{n}\simeq 0.5$), the classical bound
is raised to $68.8\%$ (see Supplementary Note 3 and \cite{23a,25,34}). Our measured
conditional fidelities $\overline{F}$ for those $105$ pairs of memory cells are above or
around $90\%$. The average of the conditional fidelities over the $105$ pairs is $(94.45\pm0.06)\%$. The standard deviations of these
measurements are shown in the Supplementary Note 4. The measured conditional fidelities
for all the memory cells significantly exceed the classical bound by more than four standard deviations.

\subsection{Characterization of storage efficiencies and efficiency-dependent classical bounds}

We then measure the efficiency of the photon storage in each memory cell.
The measurement is done by directing the weak coherent pulse (with $\bar{n}%
\simeq 0.5$) to each memory cell and then detect the probability of the
stored photon going to the output single-mode fiber after a storage time of $%
1.38$ $\mu $s. The detection is scanned over all the memory cells by
controlling the optical paths with the set of input and output AODs.
The results are shown in Fig. 3a. The efficiency ranges from $18\%$ for the
middle memory cells to about $2\%$ for the edge memory cells. The major
contributor to this inefficiency is the limited optical depth of
the atomic cloud, which is about $5$ at the center of the array and decreases
to below $1$ at the edge. According to theory, the intrinsic efficiency can be significantly
improved with moderate increase of the optical depth \cite{35}. Very recent experiments have demonstrated
impressively high intrinsic efficiencies for both strong classical pulse \cite{31} and weak coherent pulse \cite{32}.
This is achieved by a significant increase of the optical depth of the atomic ensemble through use of elongated 2D (tow-dimensional) MOT or compressed MOT. As the 2D or compressed MOT has a small cross section, it is not straightforward to extend the techniques in those experiment to allow the spatial multiplexing for realization of multi-cell quantum memories.
However, those experiments \cite{31,32}, together with the theoretical calculation in Ref. \cite{35}, demonstrates that
a large improvement in the intrinsic efficiency is possible by a reasonable increase of the optical depth of the atomic cloud.
To have a larger optical depth and at the same time a larger cross section for spatial multiplexing, one way is to prepare a larger MOT by
loading of pre-cooled atoms into the memory MOT, using the double MOT structure or an additional Zeeman slower. Alternatively, we can also try to decrease the waist diameters of the control/probe beams so that each memory cell takes a smaller cross section in the whole atomic ensemble. Eventually, it would be desirable to load the atoms into 2D arrays of far-off-resonance optical traps to increase the memory time as well as to make the optical depth more homogeneous for all the memory cells.

When we take into account the contribution of the inefficiency of the
quantum storage, the classical bound for the conditional storage fidelity will be
increased \cite{23a}. In Fig. 3b, we show the calculated classical fidelity
bound for each pair of memory cells (see Supplementary Note 3), taking into
account the contributions of both the multi-photon components in a weak
coherent pulse and the measured inefficiencies for the corresponding cells.
Our measured storage fidelities for all the $105$ pairs of memory cells
shown in Fig. 2c are still higher than the corresponding classical bounds.
To compare, in Fig. 3c we show the difference in values between the measured
conditional fidelity and the corresponding classical bounds. All the values are
positive, and all of them exceed the classical bound by at least four standard deviations. The minimum difference is $6.4\%$ here, about $4$ standard deviations above the classical bound. This confirms that we have demonstrated quantum storage for all the $105$ qubits in this multi-cell memory after taking into account of the experimental imperfections.

\begin{figure}[ptb]
\includegraphics[width=18cm]{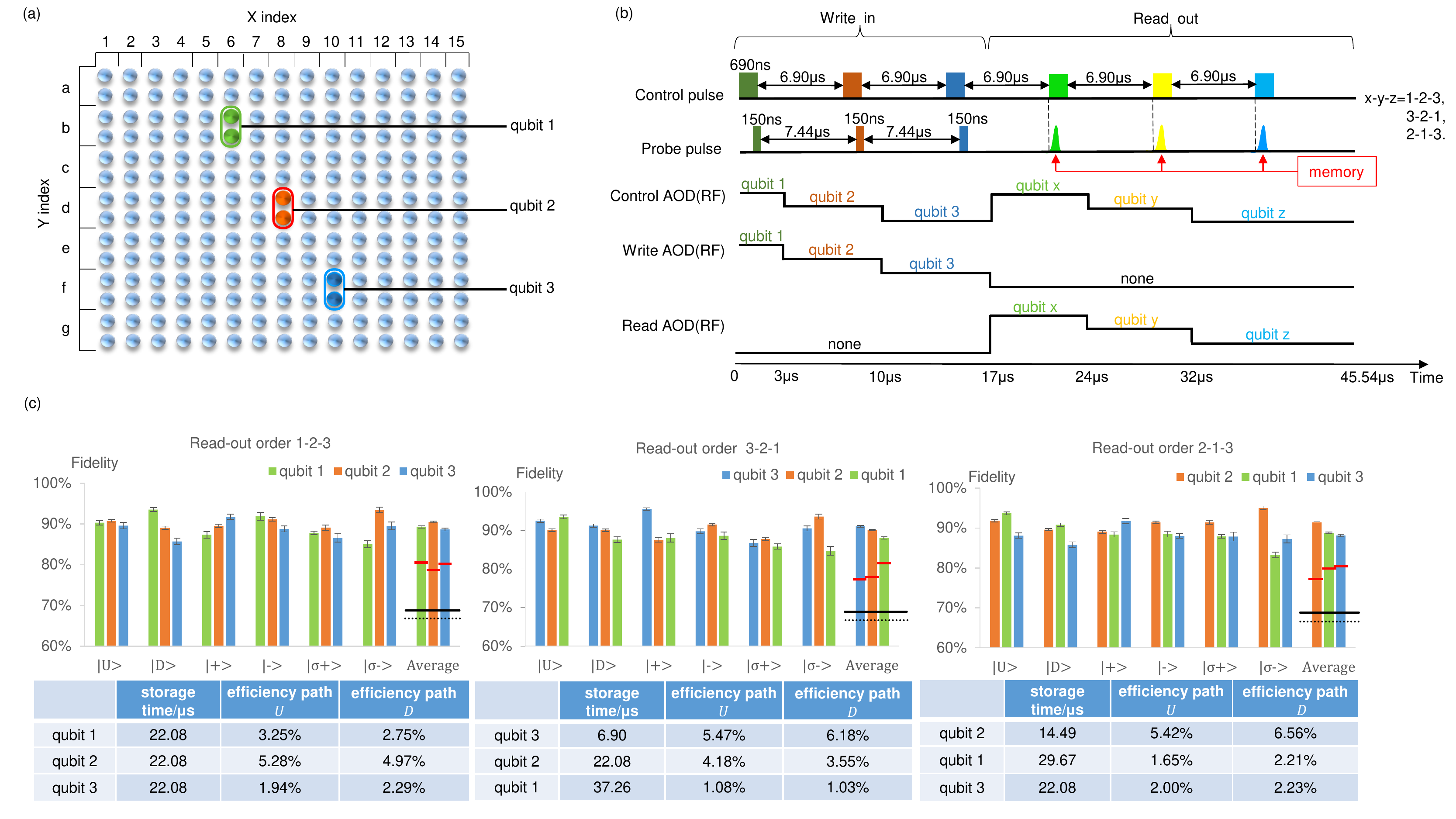}
\caption{\textbf{Demonstration of the random access feature for the multi-cell quantum memory.} \textbf{%
(a)} The positions of the three pairs of memory cells in the 2D array that we choose in this demonstration.
\textbf{(b)} The time sequences for the control pulse, the probe pulse, the radio-frequency (RF) signals that
drive the control AODs, the write AODs, and the read AODs. Different heights of the RF signals mean that
the corresponding AODs choose different pairs of memory cells for individual addressing and control. \textbf{(c)} The measured state fidelities of the three qubits stored into the corresponding atomic memory cells and read out by different orders after individually
controllable storage times. The qubits are written in by the order
1-2-3 and read out by three different orders 1-2-3, 3-2-1, and 2-1-3. The error bars denote one standard deviation of the corresponding measured values. The solid (dashed) black lines on the measured average fidelities denote the classical bounds on the fidelity with (without) the contribution of the multi-photon component in the input weak coherent pulse. The red solid lines denote the corresponding classical bound by further taking into the contribution of the limited retrieval efficiency.}
\end{figure}

\subsection{Demonstration of random access quantum storage}

Now we demonstrate the random access feature of this quantum memory. By
programming the AODs to control the optical paths, we can write multiple
photonic qubits into any of those memory cells and read them out later on-demand
by an arbitrary order. To experimentally verify this, we store three qubits
into three pairs of memory cells shown in Fig. 4a, in the order of qubits
1-2-3. After a controllable storage time, we read out these qubits in a
programmable way by three different orders. The control sequences for the
write-in and readout process are shown in Fig. 4b and can be fully
programed. In Fig. 4c, we show the storage fidelity measured through quantum
state tomography, the retrieval efficiency, and the storage time for each
qubit, with three different readout orders of qubits 1-2-3, 3-2-1, and
2-1-3. All the fidelities exceed the corresponding classical bounds, even
after taking into account of the multi-photon components and the storage
inefficiencies. The above control methods for three qubits can be similarly
applied for simultaneous storage of more qubits and programing of their
readout patterns. The current experiment is mainly limited by the memory
time in the atomic ensemble, which is about $27.8$ $\mu s$ ($1/e$ decay time), caused by the
thermal motion of the atomic gas and the remaining small magnetic field
gradient. The memory time in the atomic ensemble can be extended by orders
of magnitude if we make use of a far-of-resonant optical trap to confine the
atoms \cite{18,19}.

\section{Discussion}

Our experiment realizes a multiplexed random access quantum memory with $210$
memory cells that can store $105$ optical qubits. We demonstrate
programmable storage and readout of individual qubits in the memory cells
with the access time and the readout order fully controllable and
independent of the physical location of the cells.
This random access feature, together with the large capacity for storing optical qubits in a
single macroscopic ensemble, opens up an interesting perspective for
applications. For instance, it could be useful for realization of
multiplexed quantum repeater networks towards long-distance communication
and quantum internet \cite{1,2,3,4,5,6,7} or for optical quantum
information processing and computation that requires the memory components
\cite{8,9,10}. The random access control technique may also find
applications in other quantum information systems that require individual
addressing and programmable control \cite{27}.

In this experiment, we focus on the proof-of-principle demonstration, where the input state, although remaining unknown to the
atomic quantum memory, is prepared by the write AODs in the same experimental setup. For future applications, it would be important to store the input state free-propagating from the outside into this random access quantum memory as well. Depending on the type of the optical qubits coming from the outside, this could be done by the AODs or other linear optical devices. For instance, if the input qubit is carried by the time bins of optical pulses, we can program the AODs to split different time-bins into different optical paths in our dual-rail representation so that they can be stored into the corresponding multiplexed atomic memory cells. If the input state is carried by different polarizations of an optical pulse, we can first use a polarizing beam splitter to split them into different paths and then use AODs in those paths for spatial multiplexing. Preparation of different types of optical qubits from the outside, possibly by another atomic ensemble, and demonstration of their storage and control inside the random access quantum memory is an interesting future direction.

\textbf{Data Availability} The data that support the findings of this study are available
from the authors upon request.

\textbf{Acknowledgements:}
This work was supported by by the Ministry of Education of China
and the National key Research and Development Program of China (2016YFA0301902).

\textbf{Competing interests:} The authors declare that there are no competing interests.

\textbf{Author Information:} Correspondence and requests for materials should be addressed to L.M.D.
(lmduan@tsinghua.edu.cn).

\textbf{Author Contributions:} L.M.D. conceived the experiment and supervised the
project. N.J., Y.F.P., W.C., C.L., S.Z. carried out the experiment. N. J. and Y. F. P. contribute
equally to this experiment. L.M.D., N.J., Y.F.P. wrote the manuscript.

\newpage

\section{ Supplementary Note}

\section{ Supplementary Note 1: Experimental setup}

A $^{87}$Rb atomic cloud is loaded into a magneto-optical trap (MOT) inside
a vacuum glass cell. A strong cooling beam red detuned to the D2 cycling
transition $|g\rangle \equiv |5S_{1/2},F=2\rangle \rightarrow
|5P_{3/2},F^{\prime }=3\rangle $ by $13$ MHz is used for cooling and
trapping of the atoms. Some atoms could fall out of the cooling cycle, so we
use the repumping laser, resonant to the $|s\rangle \equiv
|5S_{1/2},F=1\rangle \rightarrow |5P_{3/2},F^{\prime }=2\rangle $ transition
to pump them back. The diameter of the cloud in the MOT is about $3$ mm and
the temperature is about $300$ $\mu $K. The atoms are then further cooled by
polarization gradient cooling (PGC) for $2$ ms. To implement the PGC, we
shut off the magnetic gradient coil, increase the red detuning of the
cooling laser to $30$ MHz, and keep its intensity the same as the value
during the MOT loading stage. At the same time, the repumping intensity is
decreased to $0.5\%$ of the value at the loading phase. The temperature is
then reduced to about $50$ $\mu $K after the PGC and the size of the MOT
remains almost the same. After the atomic state preparation, the optical
depth (OD) for the resonance to the $|s\rangle \rightarrow |e\rangle \equiv
|5P_{1/2},F^{\prime }=2\rangle $ transition reaches a value about $5$ at the
center of the cloud. During the storage, the ambient magnetic field is not
compensated, so the retrieval efficiency of the collective spin-wave
excitation undergoes the Larmor precession. The data in Fig.2 and 3 are
taken at the time set to the period of the Larmor oscillation. For Fig. 4,
the time intervals between the write and the read probe pluses are set to
integer multiples of this Larmor period to achieve the highest retrieval
efficiency.\newline

We use the electromagnetic-field induced transparency (EIT) scheme to
perform the write and the read operations in quantum memory, which convert a
probe photon into a collective spin-wave mode at the atomic ground-state
manifold by shutting off the control field \cite%
{{s1},{s2}}. The control beam
on the $|g\rangle \equiv |5S_{1/2},F=2\rangle \rightarrow |e\rangle $ has
the same circular polarization as the probe beam. The waist diameters of the
control beam and the probe beam are $135$ $\mu $m and $70$ $\mu $m,
respectively, focused on the ensemble and propagating in the forward
direction with an angle about $3^{o}$ between the control and the probe
beams. After a controllable storage time, we can retrieve the spin-wave
excitation stored in the collective atomic mode by turning on the control
field. The retrieved photon is coupled into a single mode fiber for
detection. We insert a Fabry-Perot cavity (etalon) in the path of the probe
beam before detection to filter out the diffracted control pulse.

\section{Supplementary Note 2: Multiplexing and demultiplexing of optical
circuits}

Crossed acoustic-optical defectors (AODs) are used in this experiment to
control the deflection angles of the laser beam in orthogonal directions by
adjusting the radio-frequency (RF) signal inputs to the AODs. We place a
pair of AODs in the paths of the control and the write/read probe beams to
implement the multiplexing and de-multiplexing optical circuits. Two lens
with $20$ mm focal length are inserted at the middle point between the
atomic ensemble and the write/read AODs, arranged into a $4f$-configuration
(see Fig.1 of the main text. The distance between the atomic ensemble and
each side AODs is $2f$, where $f$ is the lens focal length). With the $4f$%
-configuration, the laser beams deflected to different angles by the AODs
are focused on different corresponding positions of the atomic cloud,
enabling individual addressing of each micro-ensemble. The retrieved photon
has components coming along different optical paths with stabilized phase
difference. We fine tune the directions and positions of the pair of AODs to
maintain good interference between different paths for detection in the
superposition bases. Only multiplexing AODs are required in the path of the
control beam, with no need of de-multiplexing, as shown in Fig. 1 of the
main text. The phase differences between different optical paths are
intrinsically stable because the deflected beams go through the same
apparatus.\newline

No matter which memory cell is addressed in the atomic cloud, the probe beam
is required to be coupled into the single-mode fiber at the other side of
the atoms for detection of the transmission through all the optical
elements. In our experiment, through fine adjustment of the directions and
positions of the pair of AODs and lens, we achieve over $65\%$ coupling
efficiency for all the $210$ optical paths addressing the corresponding
memory cells.\newline

We generate the radio-frequency (RF) signal inputs to the AODs by two
arbitrary waveform generators (AWG, Tektronix 5014C). We use three channels
of each AWG in our experiment. One of the AWG supplies the RF signals for
the control, the write, and the read AODs in the X direction, and the other
supplies the RF signals of these AODs in the Y direction. The phase
differences between the three analog channels of each AWG are precisely
controlled. To address one micro-ensemble, the control beam, and the write
and the read probe beams need all to be pointed to the same position of the
atomic cloud by adjusting the corresponding RF signal frequencies to control
the deflection angles. To address different micro-ensembles, we then scan
the frequency of the RF signals for the crossed AODs from $98.8$ MHz to $%
107.2$ MHz in step of $0.6$ MHz in the X direction, and from $98.8$ MHz to $%
106.6$ MHz in step of $0.6$ MHz in the Y direction, with the deflected beams
pointing to the $15\times 14$ atomic memory cells. The retrieved photons are
collected by the de-multiplexing circuit and then coupled into the single
mode fiber for detection by a single-photon counter. By programming the AWG
to generate arbitrary RF electric signals and their superpositions for the
AOD inputs, we can direct the deflected light beams to any paths or their
superpositions \cite{s3}. \newline

\section{Supplementary Note 3: Classical bound for state storage under
experimental imperfections}

To confirm genuine quantum storage, we need to demonstrate that the fidelity
achieved in the experiment is higher than the classical bound. The classical
bound is defined as the best fidelity achievable for the same input pulse if
we replace the quantum memory by any classical means. If the input is a
single-photon pulse and the input qubit state is uniformly taken from the
whole Bloch sphere, the classical bound for the qubit storage fidelity is
known to be $2/3$ \cite{s4}. In experiments, however, it is difficult to
uniformly sample all the possible input qubit states over the whole Bloch
sphere. In our experiment, we actually only sample over six complementary
input states defined in the main text with equal weights. First, we give an
explicit elementary proof that in this case the classical bound is still $%
2/3 $. Then, we take into account the experimental imperfections, in
particular the multiphoton components in the input weak coherent pulse and
the limited efficiency of the quantum memory, and calculate the classical
bound taking into account of all these imperfections. The measured
fidelities in the main text are compared with the classical bound that
includes contributions of the experimental imperfections to confirm genuine
quantum storage.

First, let us explicitly calculate the classical bound under six
complementary input states. Assume the best measurement bases are given by $%
|\psi _{+}\rangle =\cos (\theta /2)|U\rangle +e^{i\varphi }\sin (\theta
/2)|D\rangle $, $|\psi _{-}\rangle =-\sin (\theta /2)|U\rangle +e^{i\varphi
}\cos (\theta /2)|D\rangle $, where $\theta ,\varphi $ are to-be-optimized
parameters. We calculate the average fidelity over the six input states $%
|U\rangle ,|D\rangle ,|\pm \rangle ,|\sigma _{\pm }\rangle $ . For the input
state $|U\rangle $, we have a probability of $\cos ^{2}\left( \theta
/2\right) $ ($\sin ^{2}\left( \theta /2\right) $) to obtain the outcome $%
|\psi _{+}\rangle $ ($|\psi _{-}\rangle $) which has a fidelity of $\cos
^{2}\left( \theta /2\right) $ ($\sin ^{2}\left( \theta /2\right) $). The
average fidelity is therefore
\begin{equation}
\overline{F}_{U}=\cos ^{2}\frac{\theta }{2}\cos ^{2}\frac{\theta }{2}+\sin
^{2}\frac{\theta }{2}\sin ^{2}\frac{\theta }{2}.
\end{equation}%
Similarly, for the input states $|D\rangle ,|\pm \rangle ,|\sigma _{\pm
}\rangle $, the corresponding fidelities are given respectively by
\begin{equation}
\overline{F}_{D}=\sin ^{2}\frac{\theta }{2}\sin ^{2}\frac{\theta }{2}+\cos
^{2}\frac{\theta }{2}\cos ^{2}\frac{\theta }{2},
\end{equation}%
\begin{equation}
\overline{F}_{+}=\frac{1}{4}(\cos \frac{\theta }{2}+e^{i\varphi }\sin \frac{%
\theta }{2})^{2}(\cos \frac{\theta }{2}+e^{i\varphi }\sin \frac{\theta }{2}%
)^{2}+\frac{1}{4}(\cos \frac{\theta }{2}-e^{i\varphi }\sin \frac{\theta }{2}%
)^{2}(e^{i\varphi }\cos \frac{\theta }{2}-\sin \frac{\theta }{2})^{2},
\end{equation}%
\begin{equation}
\overline{F}_{-}=\frac{1}{4}(\cos \frac{\theta }{2}-e^{i\varphi }\sin \frac{%
\theta }{2})^{2}(\cos \frac{\theta }{2}-e^{i\varphi }\sin \frac{\theta }{2}%
)^{2}+\frac{1}{4}(\cos \frac{\theta }{2}+e^{i\varphi }\sin \frac{\theta }{2}%
)^{2}(-e^{i\varphi }\cos \frac{\theta }{2}-\sin \frac{\theta }{2})^{2},
\end{equation}%
\begin{equation}
\overline{F}_{\sigma _{+}}=\frac{1}{4}(\cos \frac{\theta }{2}+ie^{i\varphi
}\sin \frac{\theta }{2})^{2}(\cos \frac{\theta }{2}+ie^{i\varphi }\sin \frac{%
\theta }{2})^{2}+\frac{1}{4}(\cos \frac{\theta }{2}-ie^{i\varphi }\sin \frac{%
\theta }{2})^{2}(-ie^{i\varphi }\cos \frac{\theta }{2}+\sin \frac{\theta }{2}%
)^{2},
\end{equation}%
\begin{equation}
\overline{F}_{\sigma _{-}}=\frac{1}{4}(\cos \frac{\theta }{2}-ie^{i\varphi
}\sin \frac{\theta }{2})^{2}(\cos \frac{\theta }{2}-ie^{i\varphi }\sin \frac{%
\theta }{2})^{2}+\frac{1}{4}(\cos \frac{\theta }{2}+ie^{i\varphi }\sin \frac{%
\theta }{2})^{2}(ie^{i\varphi }\cos \frac{\theta }{2}+\sin \frac{\theta }{2}%
)^{2},
\end{equation}%
The average fidelity over these six complementary input states under equal
weights therefore becomes
\begin{equation}
\overline{F}=\frac{1}{6}(\overline{F}_{U}+\overline{F}_{D}+\overline{F}_{+}+%
\overline{F}_{-}+\overline{F}_{\sigma _{+}}+\overline{F}_{\sigma _{-}})=%
\frac{2}{3}.
\end{equation}

Now we discuss the classical bound when the input is a weak coherent pulse.
It has been known that for a state containing $N$ qubits, the best classical
strategy leads to a fidelity of $F=\frac{N+1}{N+2}$ \cite{s5}. For an
input coherent state with a mean photon number $\mu $, the number of photons
(optical qubits) $n$ satisfy the Poissonian distribution $P(\mu ,n)=e^{-\mu }%
\frac{\mu ^{n}}{n!}$. Then, the maximum achievable fidelity becomes a
weighted  sum over $n$ of the fidelity for a given $n$ where the weight is
given by the  Poissonian distribution \cite{s6}. Combining the
above two results together, the average input-output fidelity is simply the
statistical mixture of the fidelities for $n\geq 1$ photons \cite{s7}%
:
\begin{eqnarray}
F(\mu ) &=&\sum_{n=1}^{+\infty }\frac{n+1}{n+2}\frac{P(\mu ,n)}{1-P(\mu ,0)}
\\
&=&\frac{1}{1-e^{-\mu }}[\frac{1-e^{-\mu }-\mu +\mu ^{2}}{\mu ^{2}}-\frac{%
e^{-\mu }}{2}],
\end{eqnarray}%
This bound gives a value of $68.8\%$ for our experimental case with the mean
photon number $\mu \simeq 0.5$ for the input weak coherent pulse.

Finally, we need to consider the case when the retrieval efficiency $\eta $
is significantly less than $1$. In this case, the classical memory protocol
could use a more elaborate strategy to take advantage of the finite
retrieval inefficiency to achieve a higher average fidelity when the
detector successfully registers an output photon. As shown in Refs. \cite%
{s7}, the average state fidelity conditional on a successful
registration of the output photon has the expression
\begin{equation}
F(\mu ,\eta )=\frac{\frac{n_{min}+1}{n_{min}+2}\gamma +\sum_{n\geq n_{min}}%
\frac{n+1}{n+2}P(\mu ,n)}{\gamma +\sum_{n\geq n_{min}}P(\mu ,n)},
\end{equation}%
where $0<\gamma <P(\mu ,n_{min})$ is a parameter that is adjusted to mimic
the memory efficiency $\eta $, and $n_{min}$ is the cutoff photon number to
be optimized to get the highest bound $F(\mu ,\eta )$. The supplementary
information of Ref. \cite{s7} explains how to calculate the parameter
$\gamma $ in detail with a given retrieval efficiency $\eta $. The effective
classical efficiency $\eta _{C}$ is defined as the probability that the
classical device gives an output qubit,
\begin{equation*}
\eta _{C}=\frac{\gamma +\sum_{n\geq n_{min}}P(\mu ,n)}{1-P(\mu ,0)},
\end{equation*}%
Here, the $\eta _{C}$ is assumed to be the same as the retrieval efficiency $%
\eta $, and $n_{min}$ is obtained as follows \cite{s7}
\begin{equation*}
n_{min}=\underset{i}{\min }\sum_{n\geq {i+1}}P(\mu ,n)\leq (1-P(\mu ,0))\eta
.
\end{equation*}

For our experimental data on the retrieval efficiency, the corresponding
classical bounds are shown in Fig. 3b of the main text and our measured
quantum efficiencies for the memory cells still exceed the corresponding
classical bounds taking into account of these experimental imperfections.

\section{Supplementary Note 4: standard deviation of storage fidelities for every memory cells}

\begin{figure}[ptb]
\includegraphics[width=13cm]{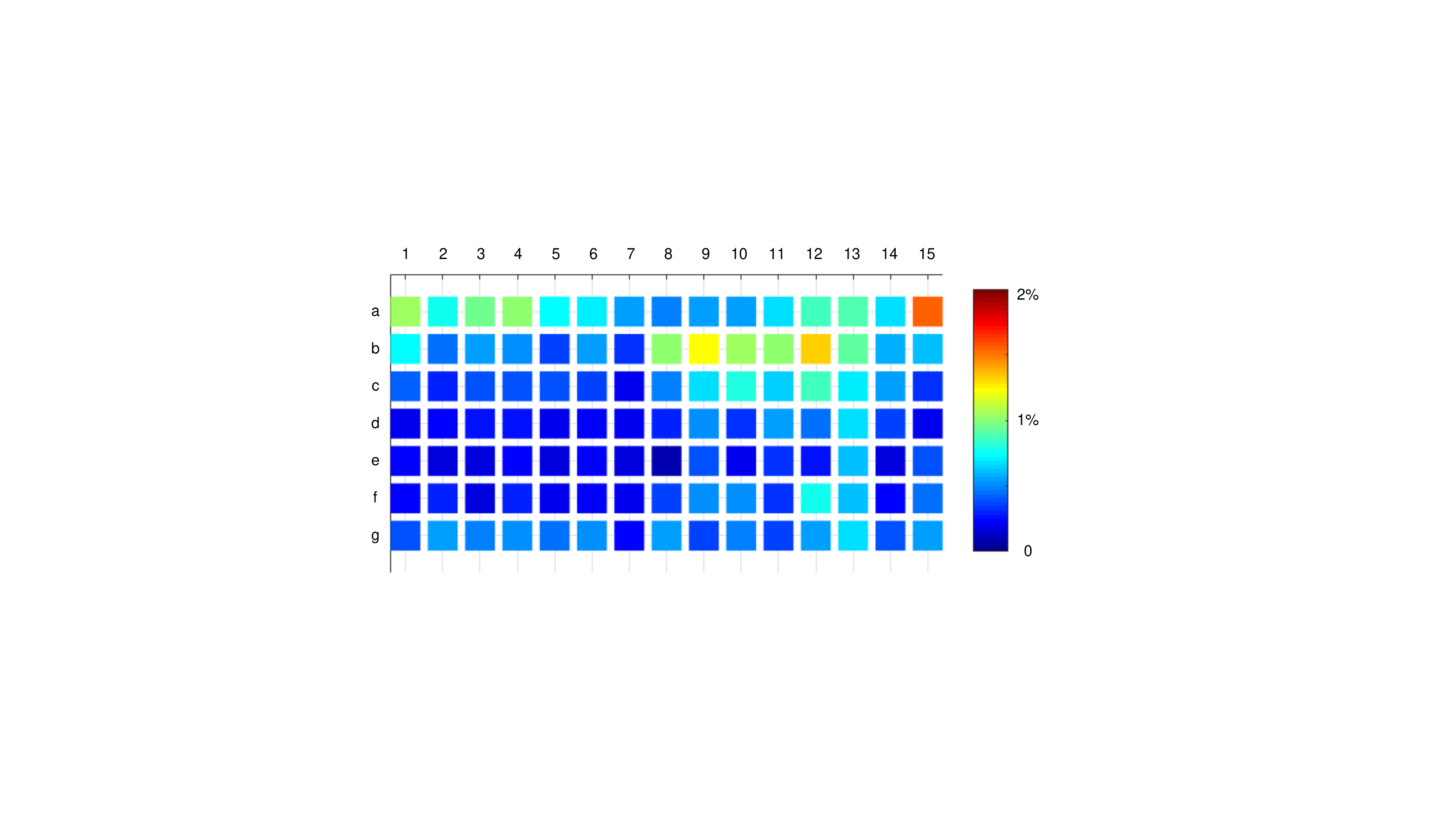}
\caption{\textbf{Measured standard deviations of the average fidelities of the retrieved optical qubits
after storage in the  $210$-cell quantum memory.}   The standard deviations of the corresponding measured average fidelities (see Fig. 2c in the maintext) for the six complementary input states of optical qubits after a $1.38$ $\protect\mu s$ storage time}
\end{figure}
For each pair of the memory cells, we measured the storage fidelity F under six complementary
input states, and the results are shown in maintext for all the 105 pairs of memory cells. Here we show the
corresponding standard deviations of the average fidelity in Fig. 5 for all the 105 pairs of memory cells. The standard deviations of averaged fidelities from the six measurements with equal weight, are  all below $2\%$.

\end{document}